\documentclass[modern]{aastex631}
\usepackage{graphicx}
\usepackage[T1]{fontenc}
\usepackage{times}
\usepackage{lineno}

\DeclareUnicodeCharacter{0301}{\'e}
\DeclareUnicodeCharacter{0308}{\'o}

\begin{document}

\title{Astrometric Calibration and Performance
of the Dark Energy Spectroscopic Instrument Focal Plane}

\correspondingauthor{Stephen Kent}
\email{skent@fnal.gov}

\author{S.~Kent}
\affiliation{Fermi National Accelerator Laboratory,
 PO Box 500,
 Batavia,
 IL 60510,
 USA}
\affiliation{Department of Astronomy and Astrophysics,
 University of Chicago,
 5640 South Ellis Avenue,
 Chicago,
 IL 60637,
 USA}
\author{E.~Neilsen}
\affiliation{Fermi National Accelerator Laboratory,
 PO Box 500,
 Batavia,
 IL 60510,
 USA}
\author{K.~Honscheid}
\affiliation{Center for Cosmology and AstroParticle Physics,
 The Ohio State University,
 191 West Woodruff Avenue,
 Columbus,
 OH 43210,
 USA}
\affiliation{Department of Physics,
 The Ohio State University,
 191 West Woodruff Avenue,
 Columbus,
 OH 43210,
 USA}
\author{D.~Rabinowitz}
\affiliation{Physics Department,
 Yale University,
 P.O. Box 208120,
 New Haven,
 CT 06511,
 USA}
\author{E.~F.~Schlafly}
\affiliation{Space Telescope Science Institute,
 3700 San Martin Drive,
 Baltimore,
 MD 21218,
 USA}
\author{J.~Guy}
\affiliation{Lawrence Berkeley National Laboratory,
 1 Cyclotron Road,
 Berkeley,
 CA 94720,
 USA}
\author{D.~Schlegel}
\affiliation{Lawrence Berkeley National Laboratory,
 1 Cyclotron Road,
 Berkeley,
 CA 94720,
 USA}
\author{J.~Garc\'ia-Bellido}
\affiliation{Instituto de F\'{\i}sica Te\'{o}rica (IFT) UAM/CSIC,
 Universidad Aut\'{o}noma de Madrid,
 Cantoblanco,
 E-28049,
 Madrid,
 Spain}
\author{T.~S.~Li}
\affiliation{Department of Astronomy \& Astrophysics,
 University of Toronto,
 Toronto,
 ON M5S 3H4,
 Canada}
\author{E.~Sanchez}
\affiliation{CIEMAT,
 Avenida Complutense 40,
 E-28040 Madrid,
 Spain}
\author{J.~Silber}
\affiliation{Lawrence Berkeley National Laboratory,
 1 Cyclotron Road,
 Berkeley,
 CA 94720,
 USA}
\author{J.~Aguilar}
\affiliation{Lawrence Berkeley National Laboratory,
 1 Cyclotron Road,
 Berkeley,
 CA 94720,
 USA}
\author{S.~Ahlen}
\affiliation{Physics Dept.,
 Boston University,
 590 Commonwealth Avenue,
 Boston,
 MA 02215,
 USA}
\author{D.~Brooks}
\affiliation{Department of Physics \& Astronomy,
 University College London,
 Gower Street,
 London,
 WC1E 6BT,
 UK}
\author{T.~Claybaugh}
\affiliation{Lawrence Berkeley National Laboratory,
 1 Cyclotron Road,
 Berkeley,
 CA 94720,
 USA}
\author{A.~de la Macorra}
\affiliation{Instituto de F\'{\i}sica,
 Universidad Nacional Aut\'{o}noma de M\'{e}xico,
  Cd. de M\'{e}xico  C.P. 04510,
  M\'{e}xico}
\author{P.~Doel}
\affiliation{Department of Physics \& Astronomy,
 University College London,
 Gower Street,
 London,
 WC1E 6BT,
 UK}
\author{D.~J.~Eisenstein}
\affiliation{Center for Astrophysics $|$ Harvard \& Smithsonian,
 60 Garden Street,
 Cambridge,
 MA 02138,
 USA}
\author{K.~Fanning}
\affiliation{The Ohio State University,
 Columbus,
 43210 OH,
 USA}
\author{A.~Font-Ribera}
\affiliation{Institut de F\'{i}sica d’Altes Energies (IFAE),
 The Barcelona Institute of Science and Technology,
 Campus UAB,
 08193 Bellaterra Barcelona,
 Spain}
\author{J.~E.~Forero-Romero}
\affiliation{Departamento de F\'isica,
 Universidad de los Andes,
 Cra. 1 No. 18A-10,
 Edificio Ip,
 CP 111711,
 Bogot\'a,
 Colombia}
\affiliation{Observatorio Astron\'omico,
 Universidad de los Andes,
 Cra. 1 No. 18A-10,
 Edificio H,
 CP 111711 Bogot\'a,
 Colombia}
\author{S.~Gontcho A Gontcho}
\affiliation{Lawrence Berkeley National Laboratory,
 1 Cyclotron Road,
 Berkeley,
 CA 94720,
 USA}
\author{J.~Jimenez}
\affiliation{Institut de F\'{i}sica d’Altes Energies (IFAE),
 The Barcelona Institute of Science and Technology,
 Campus UAB,
 08193 Bellaterra Barcelona,
 Spain}
\author{D.~Kirkby}
\affiliation{Department of Physics and Astronomy,
 University of California,
 Irvine,
 92697,
 USA}
\author{T.~Kisner}
\affiliation{Lawrence Berkeley National Laboratory,
 1 Cyclotron Road,
 Berkeley,
 CA 94720,
 USA}
\author{A.~Kremin}
\affiliation{Lawrence Berkeley National Laboratory,
 1 Cyclotron Road,
 Berkeley,
 CA 94720,
 USA}
\author{M.~Landriau}
\affiliation{Lawrence Berkeley National Laboratory,
 1 Cyclotron Road,
 Berkeley,
 CA 94720,
 USA}
\author{L.~Le~Guillou}
\affiliation{Sorbonne Universit\'{e},
 CNRS/IN2P3,
 Laboratoire de Physique Nucl\'{e}aire et de Hautes Energies (LPNHE),
 FR-75005 Paris,
 France}
\author{M.~E.~Levi}
\affiliation{Lawrence Berkeley National Laboratory,
 1 Cyclotron Road,
 Berkeley,
 CA 94720,
 USA}
\author{C.~Magneville}
\affiliation{IRFU,
 CEA,
 Universit\'{e} Paris-Saclay,
 F-91191 Gif-sur-Yvette,
 France}
\author{M.~Manera}
\affiliation{Departament de F\'{i}sica,
 Universitat Aut\`{o}noma de Barcelona,
 08193 Bellaterra (Barcelona),
 Spain}
\affiliation{Institut de F\'{i}sica d’Altes Energies (IFAE),
 The Barcelona Institute of Science and Technology,
 Campus UAB,
 08193 Bellaterra Barcelona,
 Spain}
\author{P.~Martini}
\affiliation{Center for Cosmology and AstroParticle Physics,
 The Ohio State University,
 191 West Woodruff Avenue,
 Columbus,
 OH 43210,
 USA}
\affiliation{Department of Astronomy,
 The Ohio State University,
 4055 McPherson Laboratory,
 140 W 18th Avenue,
 Columbus,
 OH 43210,
 USA}
\affiliation{The Ohio State University,
 Columbus,
 43210 OH,
 USA}
\author{A.~Meisner}
\affiliation{NSF's NOIRLab,
 950 N. Cherry Ave.,
 Tucson,
 AZ 85719,
 USA}
\author{R.~Miquel}
\affiliation{Instituci\'{o} Catalana de Recerca i Estudis Avan\c{c}ats,
 Passeig de Llu\'{\i}s Companys,
 23,
 08010 Barcelona,
 Spain}
\affiliation{Institut de F\'{i}sica d’Altes Energies (IFAE),
 The Barcelona Institute of Science and Technology,
 Campus UAB,
 08193 Bellaterra Barcelona,
 Spain}
\author{J.~Moustakas}
\affiliation{Department of Physics and Astronomy,
 Siena College,
 515 Loudon Road,
 Loudonville,
 NY 12211,
 USA}
\author{J.~Nie}
\affiliation{National Astronomical Observatories,
 Chinese Academy of Sciences,
 A20 Datun Rd.,
 Chaoyang District,
 Beijing,
 100012,
 P.R. China}
\author{N.~Palanque-Delabrouille}
\affiliation{IRFU,
 CEA,
 Universit\'{e} Paris-Saclay,
 F-91191 Gif-sur-Yvette,
 France}
\affiliation{Lawrence Berkeley National Laboratory,
 1 Cyclotron Road,
 Berkeley,
 CA 94720,
 USA}
\author{W.~J.~Percival}
\affiliation{Department of Physics and Astronomy,
 University of Waterloo,
 200 University Ave W,
 Waterloo,
 ON N2L 3G1,
 Canada}
\affiliation{Perimeter Institute for Theoretical Physics,
 31 Caroline St. North,
 Waterloo,
 ON N2L 2Y5,
 Canada}
\affiliation{Waterloo Centre for Astrophysics,
 University of Waterloo,
 200 University Ave W,
 Waterloo,
 ON N2L 3G1,
 Canada}
\author{C.~Poppett}
\affiliation{Lawrence Berkeley National Laboratory,
 1 Cyclotron Road,
 Berkeley,
 CA 94720,
 USA}
\affiliation{Space Sciences Laboratory,
 University of California,
 Berkeley,
 7 Gauss Way,
 Berkeley,
 CA  94720,
 USA}
\affiliation{University of California,
 Berkeley,
 110 Sproul Hall \#5800 Berkeley,
 CA 94720,
 USA}
\author{M.~Rezaie}
\affiliation{Department of Physics,
 Kansas State University,
 116 Cardwell Hall,
 Manhattan,
 KS 66506,
 USA}
\author{A.~J.~Ross}
\affiliation{Center for Cosmology and AstroParticle Physics,
 The Ohio State University,
 191 West Woodruff Avenue,
 Columbus,
 OH 43210,
 USA}
\affiliation{Department of Astronomy,
 The Ohio State University,
 4055 McPherson Laboratory,
 140 W 18th Avenue,
 Columbus,
 OH 43210,
 USA}
\affiliation{The Ohio State University,
 Columbus,
 43210 OH,
 USA}
\author{G.~Rossi}
\affiliation{Department of Physics and Astronomy,
 Sejong University,
 Seoul,
 143-747,
 Korea}
\author{M.~Schubnell}
\affiliation{Department of Physics,
 University of Michigan,
 Ann Arbor,
 MI 48109,
 USA}
\affiliation{University of Michigan,
 Ann Arbor,
 MI 48109,
 USA}
\author{H.~Seo}
\affiliation{Department of Physics \& Astronomy,
 Ohio University,
 Athens,
 OH 45701,
 USA}
\author{Gregory~Tarl\'{e}}
\affiliation{University of Michigan,
 Ann Arbor,
 MI 48109,
 USA}
\author{B.~A.~Weaver}
\affiliation{NSF's NOIRLab,
 950 N. Cherry Ave.,
 Tucson,
 AZ 85719,
 USA}
\author{R.~Zhou}
\affiliation{Lawrence Berkeley National Laboratory,
 1 Cyclotron Road,
 Berkeley,
 CA 94720,
 USA}
\author{Z.~Zhou}
\affiliation{National Astronomical Observatories,
 Chinese Academy of Sciences,
 A20 Datun Rd.,
 Chaoyang District,
 Beijing,
 100012,
 P.R. China}
\author{H.~Zou}
\affiliation{National Astronomical Observatories,
 Chinese Academy of Sciences,
 A20 Datun Rd.,
 Chaoyang District,
 Beijing,
 100012,
 P.R. China}

\keywords{Astronomical techniques(1684) - Wide-Field telescopes(1800) -
Calibration(2179)
}

\begin{abstract}

The Dark Energy Spectroscopic Instrument, consisting of
5020 robotic fiber positioners and associated systems on the Mayall
telescope at Kitt Peak, Arizona, is carrying out a survey to measure the
spectra of 40 million galaxies and quasars and produce the largest 3D map
of the universe to date.  The primary science goal is to use baryon acoustic
oscillations to measure the expansion
history of the universe and the time evolution of dark energy.
A key function of the online control system is to position each fiber
on a particular target in the focal plane
with an accuracy of 11$\mu$m rms 2-D.
This paper describes the set of software
programs used to perform this function along with the
methods used to validate their performance.

\end{abstract}

\section{Introduction} \label{sec:intro}

\smallskip

The Dark Energy Spectroscopic Instrument (DESI) is conducting an
optical/near-infrared survey of 40 million galaxies
and quasars in order to answer cosmological questions about the nature of
dark energy in the universe, with the goal of performing the most precise
measurement of the expansion history of the universe ever obtained
\citep{levi13,desi16a,desi16b}.
DESI will use the Baryon Acoustic Oscillation (BAO)
scale to determine the distance-redshift relationship from the
recent universe to approximately redshift 3.5. In addition to the
expansion history and dark energy, DESI will also
measure the growth of cosmic structure, provide new information on the
sum of the neutrino masses, study the scale
dependence of primordial density fluctuations from inflation,
test potential modifications to the general theory of relativity,
and map the stellar halo of the Milky Way.
The survey began on 2021 May 14 and is expected to run for 5 years.

The DESI instrument \citep{aba22}
is mounted at the prime focus of the Mayall 4-meter
telescope located on Kitt Peak in Arizona.  A 6-element optical corrector
provides a 3.2 degree diameter field of view \citep{miller23}.
An integrated atmospheric
dispersion compensator (ADC) provides correction for atmospheric chromatic
effects up to an airmass of about 2.2 (zenith distance of 63 deg).
The DESI focal plane comprises an array of 5020 optical fibers, each
mounted to a 2-axis robotic positioner,
along with ten "guide/focus assembly" (GFA) charge-coupled device (CCD) sensors
used for guiding or wavefront sensing.
\citep{silber23}.
Each fiber feeds light from an astronomical target
to one of ten three-channel spectrographs that cover the wavelength
range 0.36$\mu$m -- 0.98$\mu$m \citep{aba22}.  Light-emitting diode (LED)
light sources in
each spectrograph can be used to back-illuminate the fiber tips.
Additionally, 120 ``fiducials'' (fixed light sources) are distributed about the
focal plane to act as positional references.
A ``fiber view camera'' (FVC) \citep{baltay19} is mounted near the
vertex of the primary mirror and takes images of
the fiducials and back-illuminated
fiber tips of the focal plane through the
corrector.   These images are used to measure the location of the positioner
fibers relative to the fiducials and determine the necessary corrections
needed to position the fibers at the locations
of targets in a given field on the sky.  Spectrograph
integration times on a field
are typically 10-15 minutes.  For efficiency, short setup times (no more than
2 minutes, including telescope slew time, for fields nearby on the sky)
are desirable.

\begin{figure}[ht]
    \centering
    \includegraphics[width=4in]{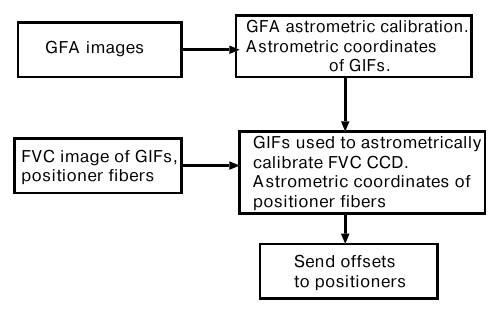}
    \caption{Flowchart showing the basic steps of the astrometric calibration
process}
\label{fig:flowchart}
\end{figure}

Accurate fiber positioning requires that the focal plane be calibrated
astrometrically.  In a nutshell, this calibration occurs in two
steps (Fig.~\ref{fig:flowchart}).
First, the GFA guide
CCDs obtain images of stars from the Gaia DR2 catalog
\citep{gaia18}, which have high-accuracy astrometric coordinates.
Two nearby co-mounted ``guider fiducials'' (GIFs),
are calibrated astrometrically, making use of previously obtained laboratory
metrology that ties CCD pixels to the GIFs.
Second, an FVC image of the fiducials and back-illuminated
positioner fibers is obtained.
Using the GIFs as surrogate astrometric standards, the FVC CCD
image is calibrated astrometrically, giving the sky position of each
pixel.  The positioners are commanded to move as appropriate to place
the image of each back-illuminated positioner
fiber on the desired pixel position of a target.

The main software program used to accomplish this calibration and
positioning of target fibers is
called ``PlateMaker'' (PM; the name is a throwback
to previous multi-object focal plane systems that used plug-plates to position
fibers - e.g., \cite{lim93}).
FVC CCD images are analyzed with a separate program called
``spotmatch,'' which derives the pixel coordinates of the fiducials and
positioner fibers.  A third program, the ``turbulence correction'' code
(part of the ``desimeter'' package) is used to measure and derive
corrections to FVC images due to the effects of turbulence in the air
column between the FVC and the focal plane.  A fourth program,
the ``dither analysis'' code, is
used offline to refine the optical model used in PM by analyzing
a special set of ``dither'' exposures of a field of bright astrometric
standard stars.
The functioning of these programs was 
described briefly in \cite{aba22}, and their integration with
the instrument control system was described in \cite{silber23}.
Modeling of the telescope optics is done using another
offline program ``trace''.

The purpose of this paper is discuss the operation and performance
of these programs,
particularly PlateMaker, in much greater depth.
The organization is as follows.  Section \ref{sec:hardware}
describes
the hardware systems that are involved and the nominal procedures by which
they are used.  Section \ref{sec:optics} gives certain details of the optical
model of the telescope and wide-field corrector.
Section \ref{sec:astrometry} describes the
astrometric calibration
of the guide CCDs used for field acquisition.
Section \ref{sec:fvc} describes the astrometric calibration of the FVC images
and how they are used to provide corrections to the fiber positioners.
Sections \ref{sec:dither} to \ref{sec:stability} present numerous measures
of the performance of the calibration procedures, including both
on-sky and off-sky tests.  Conclusions are presented in
section \ref{sec:conclusions} along with thoughts on improvements that
can be made in a future wide-field spectroscopic instrument.

\section{Hardware} \label{sec:hardware}

\begin{figure}[ht]
    \centering
    \includegraphics[width=4.7in]{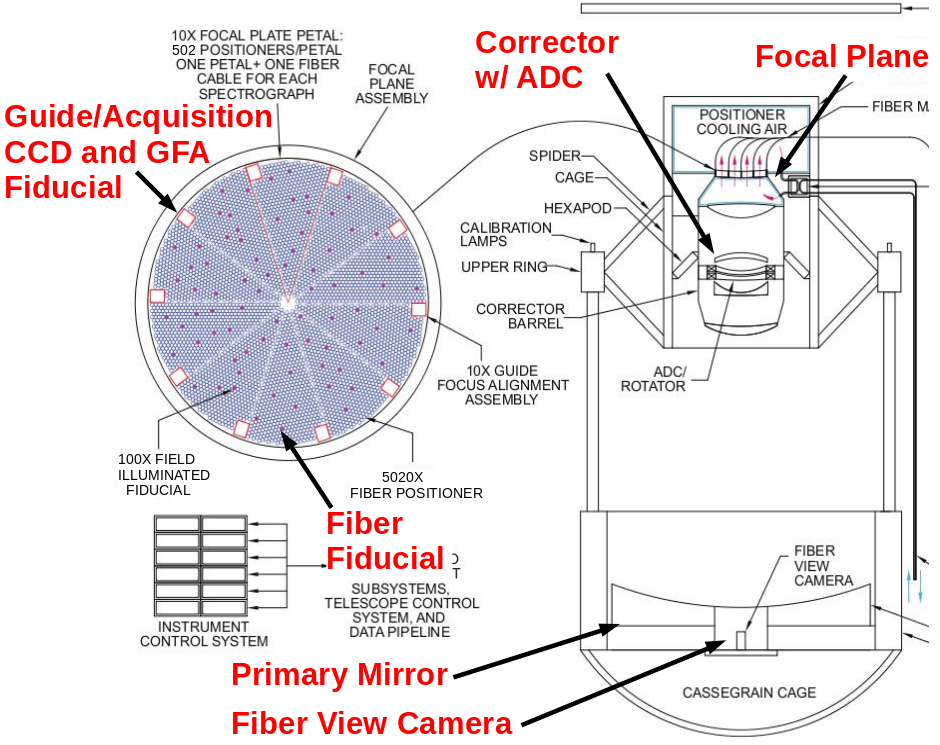}
    \caption{Schematic layout of focal plane (left) and side view of Mayall
telescope (right) showing the major DESI hardware systems relevant to
PlateMaker}
\label{fig:system}
\end{figure}

Figure \ref{fig:system} shows the basic hardware layout.  More information
on each major component follows.

\subsection{Corrector}

The basic parameters of the corrector \citep{miller23}
are give in Table \ref{tab:corrector}.

\begin{table}[ht]
\begin{center}
\caption{Optical Corrector Parameters}
\begin{tabular}{lll}
Plate Scale & 14.2 & arcsec mm$^{-1}$ \\
Field Diameter & 3.2 & deg \\
Focal Plane Diameter & 812 & mm \\
Zenith angle limits & $0-60$ & deg \\
Wavelength range & $0.36 - 0.98$ & micron \\
\label{tab:corrector}
\end{tabular}
\end{center}
\end{table}

The corrector acts as a focal expander, increasing the f/\# from 2.8 of the
primary mirror to 3.9.  The focal plane is chief-ray normal but not flat - the
shape is aspheric and convex towards the corrector.  A side effect of
producing these optical properties is that there is large amount of radial
distortion (6 mm), requiring at least a 5th order polynomial to model.
The difference between the radial and
tangential plate scales can be as large as 10\%
towards the edge of the field, which means that the exit pupil is distorted
as well.  Additionally, the atmospheric dispersion compensator
("ADC") introduces non-axisymmetric distortions;
these will be discussed in section \ref{sec:distortion}.

The corrector is attached to the telescope via a hexapod system, which 
provides focus, $x,y$ translation,
and rotational motions.  Only small rotations
are needed, and they are used to compensate for apparent field rotation as a
function of position in the sky.

\subsection{Fiber View Camera}

The FVC is described in more detail in \cite{baltay19}, so only the salient
points will be mentioned here.  The camera consists of a lens, a
narrow band filter centered on 470 nm, a
quartz window, and a Kodak KAF-50100 6132 x 8176 CCD with
6$\mu$m pixels.  The demagnification from the DESI
focal plane is a factor of 24.  Multiple lenses have been used over time;
the current lens is a BK7 singlet 25 mm in diameter and focal length 600 mm,
producing an f/24 beam.
The FVC is mounted in the central hole in the primary mirror and is about
12 m from the DESI focal plane.  Given its location near the vertex of
the primary mirror, the FVC images what is essentially the chief
ray of a back-illuminated fiber after the beam exits the corrector.

\subsection{Focal Plane}

\begin{figure}[ht]
    \centering
    \includegraphics[width=4in]{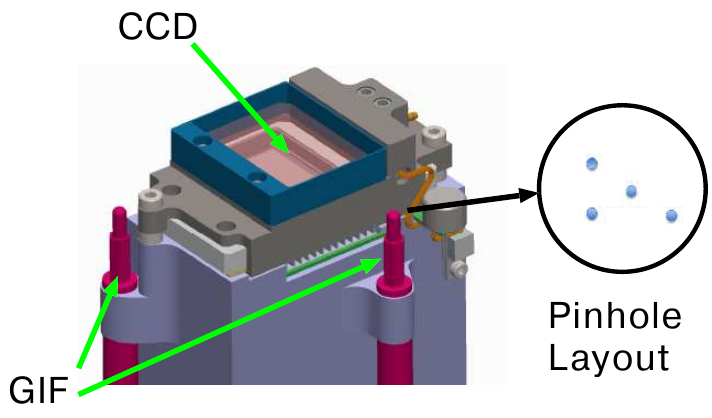}
    \caption{Close-up view of GFA assembly, showing the location of the
two GIFs attached to each assembly.  The layout of the pinhole pattern is
also shown.  {Pinholes are 10$\mu$m in diameter.}
Out of ten such assemblies around the
periphery of the DESI focal plane, six are used for guiding
and field acquisition (the others being used for focus and alignment).}
\label{fig:gfa}
\end{figure}

The focal plane \citep{silber23}
is assembled from 10 ``petals'', each being a wedged-shaped
sector of angular width 36$^{\hbox{o}}$.
Each petal has mounting holes for 502 motorized
positioners, and each positioner holds a single optical fiber of diameter
107$\mu$m (1.52 arcsec).
Additionally, 10 FIFs (``focal plane fiducials") are mounted at
select locations
between the field center and edge.  The fiducials have a mask that allows them
to project a pattern of 4 ``pinholes" when illuminated; these pinholes
are used
to aid in absolute position calibration of FVC images, tying
FVC CCD pixels to physical locations on the DESI focal plane.  Also mounted
on each petal is a single GFA assembly, consisting
of a {back-illuminated e2v Model 230-42 2048 x 2064 CCD
with 15$\mu$m pixels} plus two GIFs mounted to the GFA body
(Fig.\ \ref{fig:gfa}).  Six of these GFAs (distributed around the rim
of the focal plane) are dedicated to guiding and field acquisition, while
the remaining four are dedicated to wavefront sensing for focus and
collimation.  The GIFs are used to tie the astrometric calibration
as determined by the guide CCDs to the physical location in the focal plane.

Both the fiducials and the positioner fibers can be illuminated
either individually or in combination.
The LED illuminators have a wavelength of 470 nm.

Since the telescope has an equatorial mount, the focal plane has a
nominally fixed
orientation relative to the apparent sky.
A global Cartesian coordinate system
(called ``CS5'' within the DESI project) is defined such that the $x$-axis
points in the sky W direction and the $y$-axis points in the sky N direction.

\subsection{Metrology}

Metrology (accurate 2-D or 3-D measurements of particular components)
of two subsystems -- the GFAs and the petals -- was performed
in a laboratory setting before the subsystems were installed at the
telescope \citep{silber23}.

\begin{enumerate}
\item GFAs: Using a spot projector mounted on an X-Y stage, metrology of each
GFA was obtained, consisting of:
   \begin{enumerate}
   \item Linear measurements of the 4 pinholes of each GIF and of spots
	projected onto six locations on the CCD distributed in a grid pattern.
   \item For each spot location on the CCD, the CCD was read out,
producing pixel coordinates at the same time.  Spot sizes were about 50$\mu$m
(3.4 pixels) FWHM.
In this way, it was possible to extend the astrometric calibration of the CCD
pixel coordinates
to the GIFs.  The rms error in this measurement was of order 27$\mu$m 1-D.
   \end{enumerate}

\item Petals: After the GFAs were attached to the petals, the fiducials
(including the GIFs) were illuminated and the $x,y$ position of each
pinhole was measured relative to reference balls on each petal
using a coordinate measuring machine
(CMM) with both optical and touch probes.
The rms error in this process was of order 17$\mu$m rms 1-D.

\end{enumerate}

Originally it was intended to assemble the entire focal plane and perform
metrology of all petals simultaneously, but this proved to be too difficult
and risky, so the relative petal positions were left to be determined using
the FVC.

\section{Optical Model} \label{sec:optics}

The optical design of the corrector plus primary mirror is given in
\cite{miller23}.  This design has been analyzed using a raytrace
program\footnote{The program is home-grown, written using custom code.
It provides
many features found in commercial codes such as Zemax or Oslo but has
several enhancements that make it easier to interface to the PM code.}
to determine a number of
properties that are relevant to PM.  A somewhat unique feature of
the optical design is that the ADC is formed from two adjacent spherical
surfaces that are wedged relative to the optical axis and counter-rotated
with respect to one another in order to create lateral chromatic aberration
that compensates that due to the atmosphere.  The advantage of this
design is that it is compact and straightforward to fabricate.  A collateral
impact, however, is that numerous side effects, described below,
must be accounted for and
compensated.  None of these effects is serious, and they simply add a
few extra steps to DESI operations.

\subsection{Field Center}

The precise operational definition of the center of the focal plane
will be described in section \ref{sec:petal}; for the purpose here,
it is the place where the noses of all petals meet (Fig.\ \ref{fig:system}).
The telescope is pointed
so that a ``field center'' on the sky is positioned at this point.  The
ADCs introduce pointing offsets between the sky and the focal plane center
by an amount that depends on the ADC rotation settings and hence
zenith angle; at air mass 2, e.g., this offset is about $72''$.

\subsection{Distortion}
\label{sec:distortion}

The dominant distortion in mapping the sky to the focal plane
is 3rd and 5th order radial
distortion with amplitude of order 6 mm. (Note that this value is the maximum
amplitude relative to a linear mapping between the center and edge
of the focal plane.)  In addition, there is significant 2nd and 4th order
non-axisymmetric distortion that has a peak amplitude of order 100$\mu$m
at the edge of the field.  Both types of distortion have static
components, while the non-axisymmetric components depend additionally
on the the ADC settings.

A naive mapping of the sky to the focal plane, in which the distortion
$\Delta x$ and $\Delta y$ at coordinates $x,y$ in the focal plane
are each expressed as a two dimensional polynomial of those coordinates,
requires 20 terms total at 3rd order (e.g, Anderson and King 2003)
and 42 terms total at 5th order.  Such transformations are poorly constrained
when solving for the coefficients using data from, say, only 120 fiducials
total.  A much more efficient mapping technique was presented in
\cite{kent18} using spin-weighted Zernike polynomials.\footnote{These
polynomials turn out to be the same as the orthogonal vector polynomials
of Zhao and Burge \citep{zhao07,zhao08}(ZB) but expressed in a more easily
generalized fashion.}  These
polynomials are linear combinations of the terms in the naive expansions but
have certain properties that
make it easier to eliminate terms that are unimportant.  In practice,
only 13 terms are needed here. \cite{kent18} gives a table
of these terms along with their maximum amplitude and meaning.

\begin{figure}[ht]
    \centering
    \includegraphics[width=4.5in]{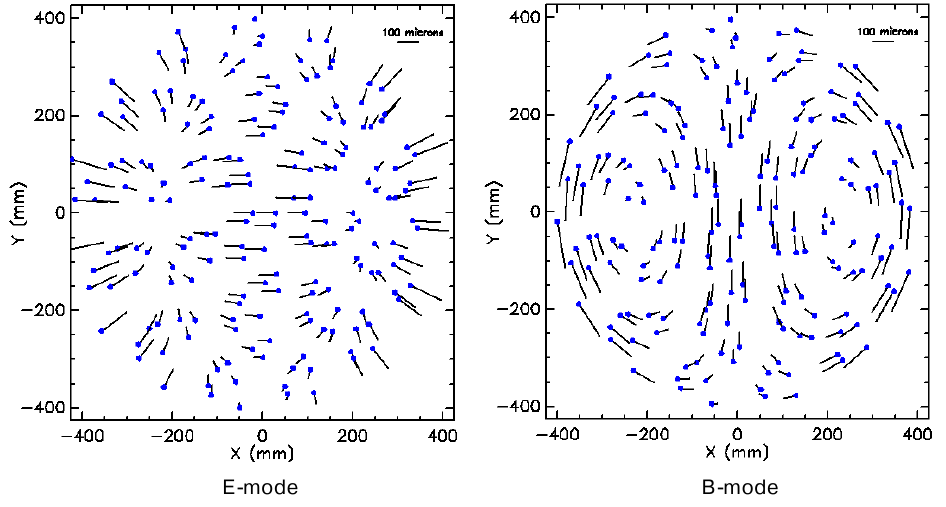}
    \caption{Distortion patterns for the two non-axisymmetric modes
induced by the ADC.}
\label{fig:ebspin}
\end{figure}

Most of the terms are type ``E'' mode while three
are type ``B'' mode.\footnote{ZB S and T polynomials respectively}
Two of these terms are non-axisymmetric modes (one E and one B) that are
generated by the ADC when the telescope is away from zenith and
the wedged lens surfaces are not parallel to one another.
The amplitudes of these terms vary sinusoidally
with the ADC angles, reaching a maximum of 112$\mu$m when the ADCs are at
their maximum correction.  Each type of pattern is shown in
Fig.\ \ref{fig:ebspin}.

When measuring distortion in the DESI focal plane using the FVC,
there is an additional
non-axisymmetric static E mode component
that is well in excess of what is predicted by the optical model and whose
origin is unclear.  A possible origin is a small misalignment of the FVC CCD
or lens with respect to the corrector optical axis.  In any case, its
effect is
incorporated in the FVC to DESI distortion model.

In the vicinity of each GFA, a simplified, local
version of the distortion model
is used that covers just the GFA CCD and the adjacent GIFs.
This model utilizes the
as-designed optical model but also accounts for the
fact that the GFAs are tilted and rotated due to the focal plane not being
flat and that the surface of best focus is curved across the GFA.
This local model omits the
non-axisymmetric E mode static term, but this omission is largely rectified
later
in the definitions of various calibration constants for the GFA CCDs and GIFs.
The local model is parametrized as a function of the ADC settings.

While the corrector plus atmosphere is largely achromatic, it is not entirely
so - there is a dependence of spot centroid on wavelength.  This dependence
is captured in a simple model consisting of $x,y$ offsets plus a radial
polynomial, where the coefficients are a function of both wavelength and
ADC settings.  {The maximum offset between the monochomatic and
polychromatic
centroids is about $13\mu$m at the GFA filter wavelength and $9\mu$m
at the FVC filter wavelength.}

The FVC images only the chief ray from a fiber, whereas the fiber itself
images all rays falling on the primary mirror.  For off-axis
images, aberrations introduce an offset between the chief
ray and the image centroid.  A simple model that is comparable to that used
for the differential chromatic aberration is used to account for this offset.
{The maximum offset between the chief ray and centroid is about $11\mu$m
at the FVC filter wavelength.}

\section{GFA Astrometric Calibration} \label{sec:astrometry}

The GFA astrometric calibration process produces a mapping from GFA CCD
pixels to sky coordinates relative to the field center.
The inputs to this PM process, as provided by the ``Instrument Control
System" (ICS), include the following:
\begin{enumerate}
\item Sky coordinates of a field of 5000 astronomical targets plus 20 blank
sky positions used for monitoring system throughput \citep{schlafly23}.
\item A list of stars from the Gaia DR2 catalog that fall
in the vicinity of each GFA CCD to serve as astrometric standards.
\item State information (hexapod settings, etc).
\item System time synchronized to UTC.
\item An image of the sky made using the GFA CCDs with the telescope
pointed at the selected field center, with a typical exposure time of
10 sec.
\end{enumerate}

\subsection{Image Processing}

The GFA images are processed to produce a list of all stellar objects
in the images.
The GFA CCDs are operated in a mode slightly different from that of most
science CCDs and thus require a somewhat nonstandard processing algorithm.
The CCDs are operated in frame transfer mode, which allows integration
and readout without the need for a shutter.
They are not actively cooled and thus have significant dark current
and hot pixels.  When the CCDs are first
turned on, they are in a very noisy state.  By running through a series
of resets
the noise levels can be reduced, however, sometimes this process
does not work.  Some of the CCDs have hot columns or warm edges.

Fortunately the hot pixel pattern has stayed nearly
constant with time such that a template dark current frame constructed at the
beginning of the survey still suffices.
After scaling for exposure time, it is subtracted
from a GFA image.  There are often gradients
in the background remaining, so the image is divided into regions of size
172x128 pixels and each is median averaged and subtracted. No flatfielding is
needed.  A simple cosmic ray filter is run to clean out obvious
single-pixel artifacts.  A smoothing filter is run once to reduce noise.
At this point a simple peak-finder is run to find local peaks greater than
a given threshold above sky noise and less than a saturation value.
This process generally finds all bright stars, but it can also find a number
of artifacts such as bad columns that need to be filtered out.  A series of
filtering steps is run to weed these out.  First, objects falling on
bad columns are identified and eliminated.  Annular profiles are computed for
each remaining object and a profile (essentially a Moffat function with
beta = 3.5) is fitted.  A fixed radius of 12 pixels is used regardless
of the seeing to avoid potential changes in profile shape during times of
poor seeing.  Cuts are imposed on the quality of the fit, image width,
flux within the profile, and shifts in centroid.  This process
handles just about any artifacts found in the images and returns a clean
list of stars.

\subsection{Global Astrometric Solution}

\begin{figure}[ht]
    \centering
    \includegraphics[width=4in]{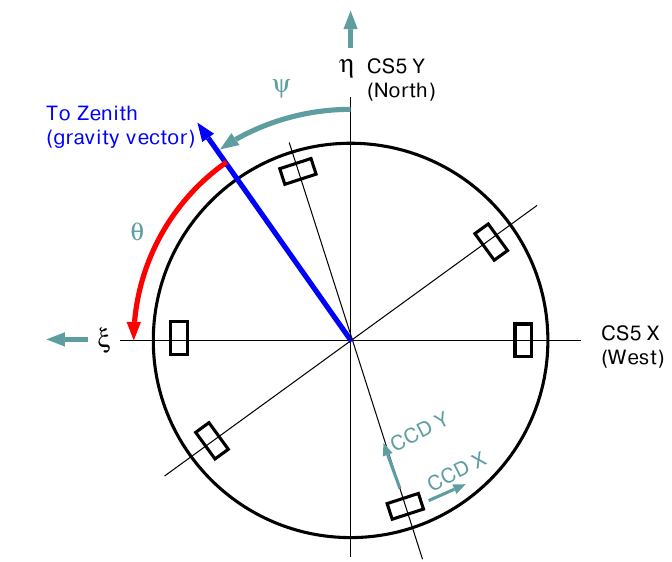}
    \caption{Layout of guider GFAs on the DESI focal plane with various
coordinate systems labeled. The parallactic angle $\psi$ is the position
angle of zenith relative to north. The position angle of a typical GFA
relative to zenith is marked as $\theta$.}
\label{fig:focal}
\end{figure}

The six GFA CCDs are treated as a single rigid focal plane - each
list of stars is projected onto the sky based on the known location and
orientation of each CCD (Fig.\ \ref{fig:focal})
and then combined into a single list.

The Gaia astrometric standard coordinates are converted to apparent
tangent plane coordinates ($\xi, \eta$) centered on the target field center
by applying standard corrections for precession,
nutation, aberration, and refraction.  An important additional correction
involves field rotation due to misalignments within the telescope structure.
Since the Mayall telescope is an equatorial mount, apparent north
is nominally fixed
in the focal plane, but due to misalignments (e.g., polar axis
misalignment), the actual direction of N can vary by several arcmin,
being largest at high declinations.  An empirical model was developed
to calculate rotation as a function of hour angle and declination, and
this model works with an rms error of 10 arcsec and peak errors of order
1 arcmin at high declination.  (These errors are not entirely
repeatable, indicating
that there may be hysteresis of some sort in the telecope drive or support
system.)

Transformation from the instrumental coordinates to sky coordinates requires
a minimum of 4 parameters: $x$ and $y$ sky center offsets, rotation angle, and
scale factor.
The rotation angle and scale factor are known well enough that the dominant
unknowns are the sky center offsets.
These offsets are determined by performing
a 2-dimensional cross-correlation between the combined star list and the
list of astrometric standards.  Once an initial guess at the offsets is known,
the observed and standard star lists are matched up in detail, and improved
values of the 4 parameters are computed via least squares.  This process
is repeated once in case the initial match missed one or more stars or
mismatched standard stars with a particular detection.  A limiting magnitude
of 18 is imposed to prevent confusion in fields with high stellar density.
Note that even if only 1 star is detected on a GFA, it can be included in
the astrometric solution.

Once the 4 parameters are determined, the following quantites are computed:
\begin{itemize}
\item Coordinates and residuals for all matched stars
\item A list rank ordered by magnitude of candidate guide stars for
each GFA with both sky and pixel coordinates
\item Pointing corrections and the hexapod rotation setting.  Note that for an
equatorial telescope, for any field off the celestial equator, a motion
in right ascension induces additional field rotation that varies as
$\tan(\delta)$; this extra rotation is included in the hexapod setting.
\item Sky coordinates for each GIF.  These are computed based on laboratory
metrology of the location of each GIF relative to GFA CCD pixels along with
optical distortion predictions from the as-built DESI optical model
(Section \ref{sec:optics}).
\end{itemize}

For fields near zenith, typical rms residuals under good conditions are
30 milliarcsec 1-D with little if any dependence on position in the focal
plane.  This performance is close to that achieved
with the Dark Energy Camera using exposures that are 3 times longer
\citep{bernstein}.
For fields away from the zenith, the accuracy declines, likely
due to deformation of the focal plane under gravity loading.  Deformation
will be discussed in section \ref{sec:deform}.

The total elapsed time, including overhead for communicating with the ICS,
is typically 4 seconds.

\subsection{Local Astrometric Calibration}

PlateMaker offers an alternative mode in which each GFA CCD is astrometrically
calibrated individually.  Although this mode is not used in normal operations,
it allows astrometric solutions to be obtained when either the telescope
has large pointing errors or the field rotation is not known ab initio.
It is most commonly used when constructing a new telescope pointing model.

\subsection{Updates to GFA metrology}
\label{sec:update}

Initially the location and orientation
of each GFA was taken from the as-designed focal plane.  In practice, the
as-built focal plane location of
each GFA is offset and rotated relative to its
design value.  By using a series of exposures taken near zenith, the
astrometric solutions were used to improve the locations and
orientations of each GFA CCD relative to the others.  Note that the
absolute locations, rotation angle relative to the focal plane, and
an overall scale factor giving the absolute separations of the CCDs was
uncontrained because the astrometric solution includes parameters that
are degenerate with these terms and will compensate for any adjustment to
them.  The absolute location, size, and orientation of the GFA array
relative to the focal plane was done using the FVC as described in
section \ref{sec:gfa2}

\subsection{Focal Plane Deformation}
\label{sec:deform}

\begin{figure}[ht]
    \centering
    \includegraphics[width=4.7in]{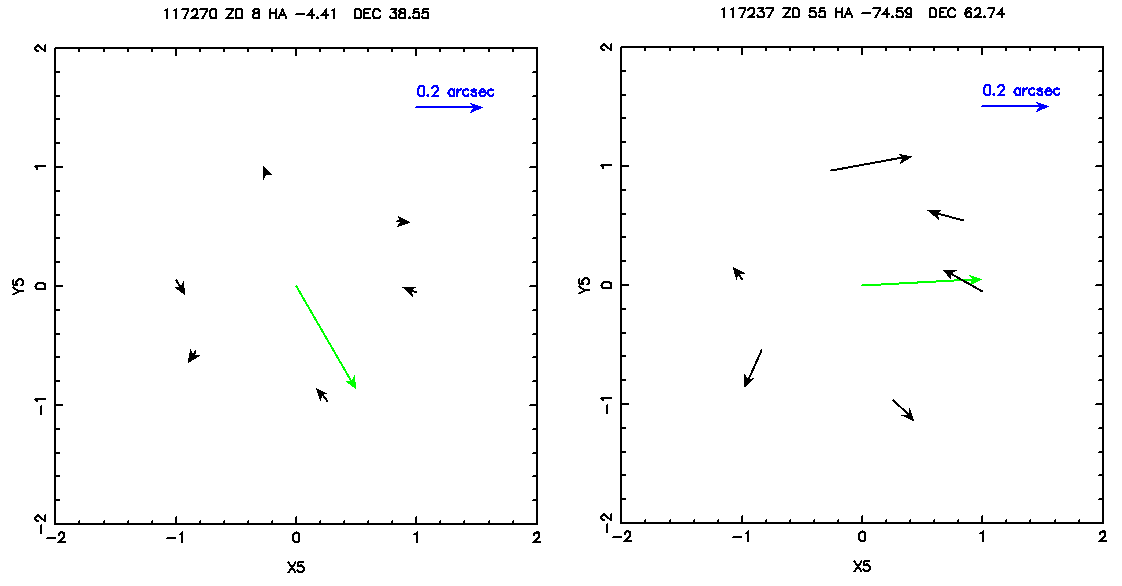}
    \caption{Left: Pattern of astrometric residuals for each GFA for a
field near zenith.  Green arrow points towards the zenith.  Right:
Same for a field at zenith distance of 55 degrees.}
\label{fig:zdcombined}
\end{figure}

\begin{figure}[ht]
    \centering
    \includegraphics[width=4in]{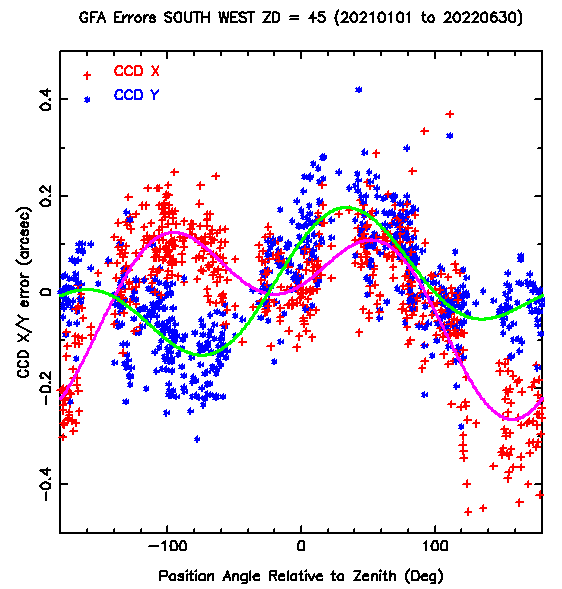}
    \caption{Astrometric residuals per GFA as a function of position angle
w.r.t. zenith (measured CCW around the DESI focal plane)
in the CCD X (red) and Y (blue) directions for a set of fields in the SW at
zenith distance 45 degrees.  Solid lines are a model fit.}
\label{fig:gfasw45}
\end{figure}

\begin{figure}[ht]
    \centering
    \includegraphics[width=4in]{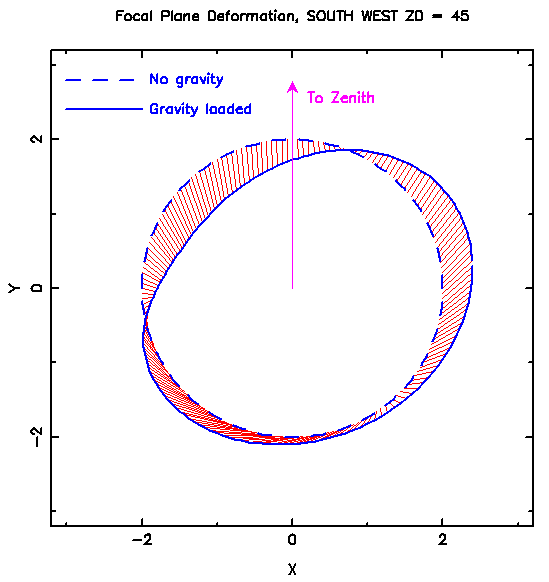}
    \caption{Exaggerated view of the focal plane deformation pattern for
the field shown at the right in Fig.\ \ref{fig:zdcombined}.  The peak
amplitude is $0.18''$.  The axis units are arbitrary.}
\label{fig:gfapat}
\end{figure}

Finite element analysis models predict that the focal plane will deform
due to gravity as the telescope moves away from the zenith \citep{miller23}.
Analysis of
astrometric solutions for a large number of fields at different locations
in the sky does, indeed, show such an effect.  Figure \ref{fig:zdcombined}
shows
the astrometric residuals averaged over each GFA for two fields: one
near the zenith, showing essentially no systematic offsets, and the other
at a zenith distance of 55 degrees, where
the residuals are large and systematic.  Figure \ref{fig:gfasw45} shows the
astrometric residuals as a function of position angle with respect to zenith
for a set fields in the SW part of the sky at a zenith distance of
45 degrees.  The large systematic trends are obvious, with a
peak amplitude of up to 0.3 arcsec.  A sinusoidal model has been fit to
each of the CCD X and Y directions with terms that are a function of both
$\theta$ and $2\theta$.  The coefficients of these terms were then fitted
as a function of zenith angle, with four different sets used for the NW, NE,
SW, and SE quadrants of the sky.  Figure \ref{fig:gfapat} illustrates
the distortion pattern for one particular field.

Corrections for deformation are applied to the GFA locations before performing
the astrometric solutions.  The model manages to reduce the impact of
deformation by about a factor of two, although not eliminating it completely.

\begin{figure}[ht]
    \centering
    \includegraphics[width=4in]{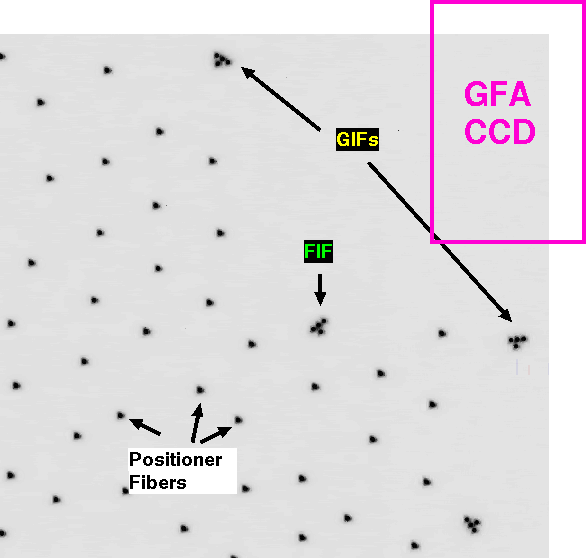}
    \caption{Portion of a FVC image of the focal plane with positioner
fibers and fiducials (FIFs and GIFs) back-illuminated.  The fiducials have a
4-spot pattern to aid in their identification.  The approximate location
of a GFA CCD is also shown.}
\label{fig:backlit}
\end{figure}

\section{FVC Astrometric Calibration} \label{sec:fvc}

Targets in a particular field are selected and assigned to specific
positioners in advance.  The layout assumes that the direction of ICRS North
has a particular position angle with respect to the focal plane CS5 Y
direction (this value being chosen
to account for precession at an epoch near the midpoint of the survey).
The target list is input to PM, which then converts the target positions
to apparent tangent plane coordinates using the same steps as were used
for the astrometric standards.  These coordinates are then converted to
focal plane locations using the as-built optical model, and the positioners
are commanded to go to these locations in an open-loop fashion.  The
petal controllers then report back the estimated location of the positioner,
accounting for any disabled or otherwise malfunctioning positioners.

The FVC takes images of the DESI focal plane with fiducials and/or positioner
fibers back-illuminated (Fig.\ \ref{fig:backlit})
and analyzes them using a software program spotmatch
(SM).
In order for SM to function it needs to be provided with a list of
approximate guesses at the locations in the image of each ``spot'' or array
of spots from each device, at which
point it is able to identify the complete pattern and return the original
list augmented with the precise pixel location of each spot.  PlateMaker
runs a separate procedure to prepare this list.  It first receives as
input from the ICS a list of all fiducials that
are operational and the list of positioners with their actual
locations.  It then converts them to pixel coordinates, making use of the
optical model of the telescope and an optical model of the FVC camera.
The predictions account for demagnification from the DESI focal plane to the
FVC CCD and rotation due to the hexapod; SM itself can account for any
overall translation of the field.  The predicted positions are generally
accurate to 1-2 pixels.

Once PM receives back the list of fiducials and positioners with pixel
coordinates, it uses the GIF coordinates, which have known
astrometric positions
from the GFA astrometric calibration, to solve for an astrometric
calibration of the FVC CCD.  Because the FVC singlet lens acts
essentially as a perfect pinhole camera, the mapping of the sky to the
CCD is a gnomonic projection, and the calibration consists of determining
a field center, scale factor, and rotation.
At this point PM can now calculate the actual sky position of each
positioner fiber, compare it with the desired position, and derive
offsets and corrections in focal plane coordinates needed to fine-tune
each positioner fiber location.  These correction moves are sent back to the
ICS and applied.  A final FVC image is taken to assess
the final set of offsets, which are recorded but not used in another
iteration.

The total elapsed time for each iteration of this process is typically
9 seconds and is the dominant contribution to the overall elapsed time
for the combined PM processes; however, the overall setup time (including
image acquisition times) is still within the 2 minute
requirement on setup time for moving among fields closely spaced on the
sky \citep{aba22}.

It should be noted that, as far as PM is concerned, absolute focal plane
locations of positioner fibers are not needed.  However, they are needed
for bookkeeping purposes and for the anti-collision code (which ensures
that positioners do not collide when moved) to work.  For this
purpose, the fiducials are used to determine a mapping from FVC pixels
to DESI focal plane coordinates using the optical model, solving
each time for the coefficients in the distortion model.

\subsection{Turbulence}

An effect that turned out to be larger than expected was
the impact of air turbulence (a.k.a. ``dome seeing'') in the 12 meter
column between the FVC and the DESI focal plane,
which introduces time-dependent
distortion in the spot position locations.  While the distortion was
expected to be of order 3$\mu$m \citep{wang14,silber23}, at times it
can amount to 10$\mu$m or more.
In fact, it is the dominant error in positioning fibers on targets,
to the extent that in the worst case the mispositioning causes
a spectroscopic exposure to be rendered useless.  Example images are
presented in \cite{silber23}.

\begin{figure}[ht]
    \centering
    \includegraphics[width=4.7in]{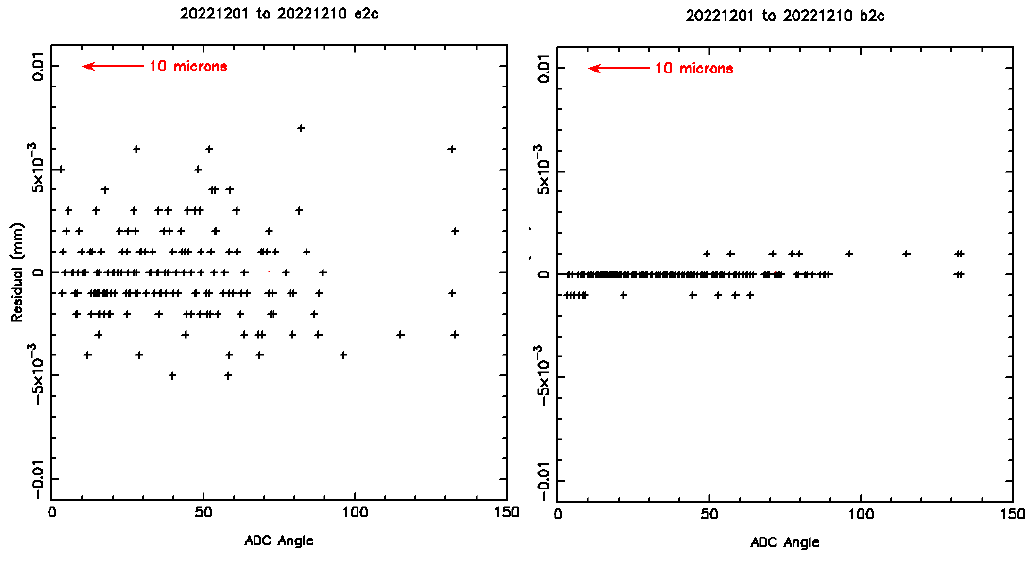}
    \caption{Amplitude of the two non-axisymmetric modes
in Fig.\ \ref{fig:ebspin} in the distortion model for a set of
210 exposures taken at various ADC settings.  The quantity plotted is
the residual between the measured coefficient and that predicted by
the telescope optical model.  Left: E-mode.
Right: B-mode.
Turbulence affects the amplitude of the former (rms
scatter of $2.4\mu$m) but
not the latter (rms scatter of $0.3\mu$m).
}
\label{fig:ebmode}
\end{figure}

Low-order modes in the turbulence pattern affect the distortion model
solutions.  In the limit that abolute variations in refractive index
are small ($\approx 10^{-6}$), the displacement of a particular
spot as recorded on the FVC CCD is given by the gradient
of the integral of air density fluctuations along the sight line
from the DESI focal plane to the FVC.  Thus, the
turbulence pattern should affect the curl-free E modes but not the
gradient-free B modes of the distortion model.  This property is
demonstrated
in Fig.\ \ref{fig:ebmode}, where the coefficients for a pair of
E-mode and B-mode terms from a set of exposures are compared, showing that
turbulence does, indeed, impact only the E-modes.  A comparable effect
was seen in the impact of atmospheric turbulence on the
astrometric calibration of the Dark Enery Camera focal
plane \citep{bernstein}.

\begin{figure}[ht]
    \centering
    \includegraphics[width=4in]{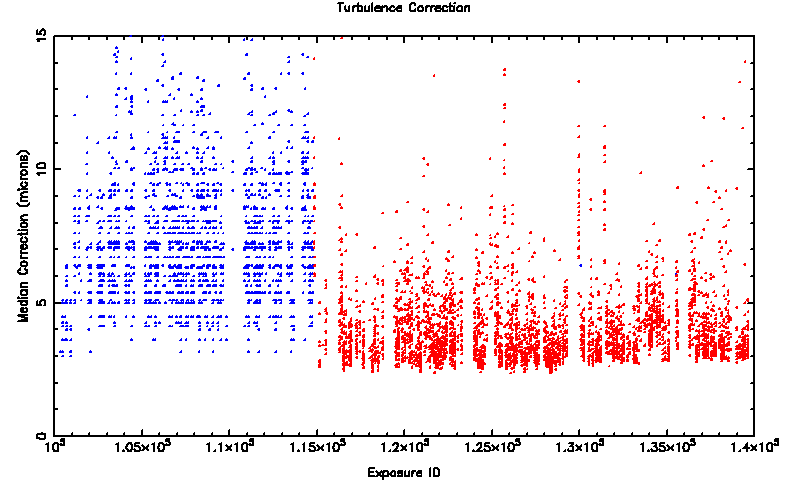}
    \caption{Median fiber positioning error over the focal plane
before (blue) and after (red) implementating the turbulence correction code.
The floor at about 3 $\mu$m is likely set by the intrinsic
rms centroiding error of a spot image as measured by
spotmatch using the FVC CCD.}
\label{fig:turbulence}
\end{figure}

One way to mitigate turbulence would be to take repeated FVC exposures to
average out the effect.  However, it was discovered that one can use the
fiducials and functional fibers in disabled positioners (together,
positional references) to calibrate and
greatly reduce the impact of turbulence.  To do this, one first needs to
determine the steady-state undistorted
locations of all the references relative to one another.
These locations turn out to depend on hour angle and declination and are
impacted by the focal plane deformation due to gravity as was described
above for the GFA CCD locations.\footnote{For reasons that are not entirely
understood, a model for focal plane deformation constructed using the
positional references and the model constructed from the astrometric
calibration residuals differ by up to 7$\mu$m,
implying that the GFA assemblies
somehow move relative to the petals on which they are mounted due
to gravity.}  Therefore, an independent model for each positional reference
was constructed by taking a series of FVC images with the telescope
driven to a selected set of hour angle and declination settings.  For these
images, the conversion from FVC CCD pixels to focal plane coordinates was
performed using the as-designed optical model since the normal transformation
derived by PM, which updates the model coefficents, removes some of the
turbulence pattern that one is trying to calibrate.  Given these
steady-state locations, the first FVC image obtained in normal operations
(after the open loop positioner moves) is analyzed to determine the
turbulence pattern, which is then removed before computing the correction
moves.  The process is then repeated for the second FVC image after the
correction moves are applied.  The residual errors in positioner locations
after the correction move are considerably reduced by this turbulence
correction, and median 2-D values are typically of order 4$\mu$m
(Fig.~\ref{fig:turbulence}).

\subsection{Lens Polishing and Homogeneity Errors}

The FVC images only a small portion of the beam from a back-illuminated
fiber, centered on the chief
ray.  While the astrometric calibration procedure
accounts for any offsets of the chief ray from the centroid of the imaged
beam due to the as-designed corrector,
it does not account for any offsets introduced by imperfections in
the corrector optics, and in particular those due to polishing errors on the
lens surfaces and
inhomogeneities in the glass refractive indices.  It can be noted that
\cite{wang14} found it
necessary to outfit the Subaru telescope metrology camera with a 110 mm
aperture lens in order to reduce the impact of high
spatial frequency polishing errors in the Subaru prime focus corrector.

Detailed requirements
on polishing errors,
divided into low, medium and high spatial frequency ranges,
were specified for each surface of each lens at the time of fabrication
\citep{miller23}.  The values actually achieved were roughly a factor 2
better than the requirements.  A detailed analysis suggests that these errors
will introduce offsets in the chief ray relative to the centroid of order
3$\mu$m rms 1-D.  A related analysis based on measurements of the glass
inhomogeneities suggests an additional contribution from this
cause of about 2$\mu$m for lenses C1-C4 and the ADCs.

One feature that has proven to be more problematic is a ``divot" introduced
by a machining error on the aspheric surface of lens C3.  This feature
has a diameter of approximately 50 mm and a depth of about 1$\mu$m.
Since the FVC beam size is only 1.8 mm at this surface (compared to the
overall beam size of 256 mm),
the FVC positions of about 35-40 fibers are impacted, introducing larger
positioning errors and reduced throughput.
So far the errors introduced by this feature remain uncorrected.

\subsection{Updates to GFA Metrology - II.}
\label{sec:gfa2}

It was mentioned in section \ref{sec:update}
that the positions of the GFAs relative to
one another can be measured quite accurately,
but that there is still an ambiguity in overall scale factor, rotation angle,
and offset relative to focal plane coordinates because these cannot be
determined from astrometry alone.  However, the GIFs, which are calibrated
both astrometrically and via laboratory metrology, provide an absolute
tie to focal plane coordinates, and the FVC images are used to provide
these missing calibration constants.

\subsection{Updates to GIF Metrology}

The GIFs are used to transfer the astrometric calibration from the GFA
CCDs to the FVC CCD.  Due to errors in the GIF metrology, there are
still systematic residuals in the GIF positions relative to one another
as is seen in their residuals from the FVC CCD astrometric calibration.
The GIF metrology was revised so that the relative GIF positions are
placed on a self-consistent system while the overall astrometric
calibration was unchanged.  It can be noted that the corrections to the
relative GIF positions, when transformed astrometrically to the focal plane,
were also measured by the FVC images; these latter measurements were left to
provide a cross-check on the first set of revisions.

The internally consistent GIF metrology could still have an overall
external systematic offset
in location relative to the GFA CCDs that would not be captured by
either the astrometric calibrations or the FVC images.  Any such offset
will be dealt with in the next section.

\subsection{Split Exposures}

For fields requiring long exposures, changes in differential refraction
due to changes in parallactic angle and/or zenith distance
can cause targets to become misaligned with their fibers, by up to half
an arcsecond in the worst cases.  Rather than
conduct a complete new field acquisition, a mode has been implemented
that predicts the changes in fiber positions due to changes in
refraction and to any new ADC setting that is required
and determines the relative
adjustments needed to each fiber positioner and to the guide stars.
These adjustments
are applied open loop.  The largest uncorrected effect turns out
to be that of field rotation.  At present it is simply monitored after that
fact using the guide star measurements, although active feedback is
planned for the future.

\section{On-Sky Dither Tests}\label{sec:dither}

While the stellar astrometric calibrations are quite accurate internally
(rms errors much smaller than a fiber diameter),
the tie of the focal plane to the sky
still has an unknown uncertainty
due to the indirect manner in which the tie was
established. The tie could not be verified directly due to the lack
of any positioners with coherent imaging capability.
Thus, it was found necessary to conduct additional
tests to better measure any residual errors in the transformations and,
if necessary, apply corrections.

The technique that was found to work best \citep{aba22}
was to observe a field of
bright stars and take a set of 13 exposures, first with the fibers placed
at the nominal positions of the stars, and then with random offsets
(``dithers'') applied in each of $x$ and $y$, with each offset drawn from a
Gaussian distribution of 1-$\sigma = 0.7{''}$.
The throughput of each star was determined from spectroscopic
exposures in each of the B, R, and Z channels.  The telescope was actively
guided during each exposure.  In essence, what the dithering accomplished
was to measure the throughput of a star at a grid of locations offset
from the nominal target position,
allowing the determination of the offset that gave the maximum throughput.
In practice such a test performed on a single star would suffer from the
effects of misguiding, seeing variation, and transparency variation.
By observing a large number of stars with positions that were dithered
independently of one another,
these effects could be disentangled from each other and from
those due to individual positioner offsets.

\begin{figure}[ht]
    \centering
    \includegraphics[width=4in]{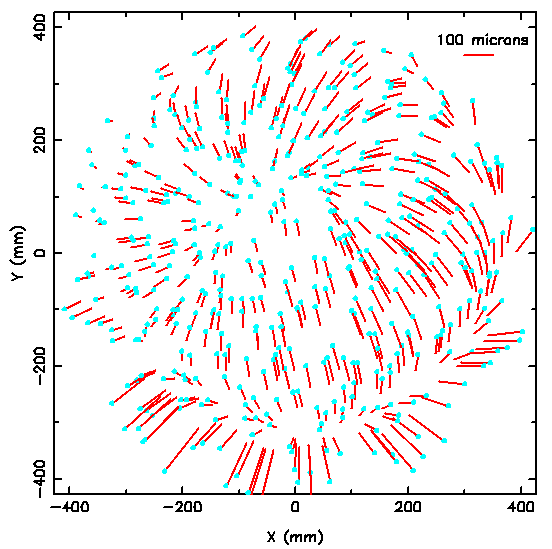}
    \caption{Quiver pattern as a function of focal plane coordinates
used to correct positioner fibers for residual
uncorrected distortion in the PM model.}
\label{fig:quiver}
\end{figure}

Having measured the residual distortion in the PM optical model, the ``quiver''
pattern (Fig.\ \ref{fig:quiver})
was fit with the same type of model as is used to map the FVC CCD to
the focal plane.  The largest term is an E mode with amplitude 56$\mu$m;
there is also a B mode term with amplitude of 14$\mu$m.  While the
origin of these terms is not known for sure, it is equivalent to a
tilt of the FVC CCD with respect to the DESI focal plane
of about 25 arcminutes.

\begin{figure}[ht]
    \centering
    \includegraphics[width=4in]{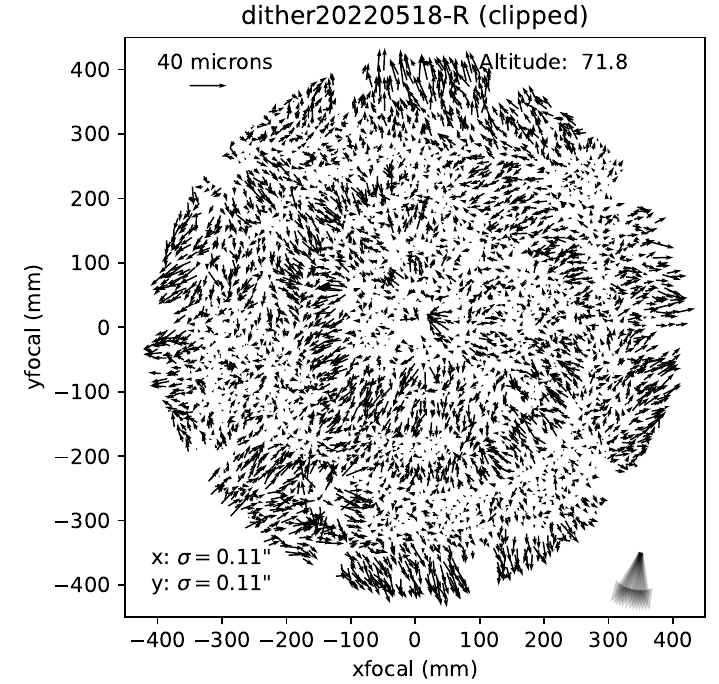}
    \caption{Residual positioner offsets in a typical dither sequence
as a function of focal plane coordinates.
(Note that $0.11''$ corresponds to 7.7$\mu$m.)}
\label{fig:residuals}
\end{figure}

The pattern made with the R channel data
is applied as a correction to the predicted target position by PM.
Subsequent dither tests show that the residual rms 1-D positioning
error is of order 8$\mu$m (Fig.~\ref{fig:residuals}).  The known
contributions from
lens imperfections (3 $\mu$m from polishing errors, 2$\mu$m from glass
inhomogeneities), and uncorrected dome turbulence plus spot centroiding
errors (4$\mu$m)
account for perhaps 5$\mu$m or 6$\mu$m of this error.
(For reference, the original allocation
to overall fiber lateral positioning error was 7.8$\mu$m 1-D or 11$\mu$m 2D,
which means that this requirement is, indeed, being met.)

The quiver pattern exhibits some high order behavior that is not included
in the model and has the rough appearance of what might be expected from
polishing errors and/or glass inhomogeneities
in the corrector optics, although this conjecture has not been verified.
Overall, the pattern and the overall centering possibly
show some dependencies on time and zenith angle, but these dependences
are at the threshold of detection and are not yet well quantified.

A byproduct of the dither tests is that it is possible to detect any
asymmetry in the geometry of the FVC CCD, particularly a difference in
scale factor in the row and column directions. In the event,
no such asymmetry is seen at the level of a few ppm.

\begin{figure}[ht]
    \centering
    \includegraphics[width=4in]{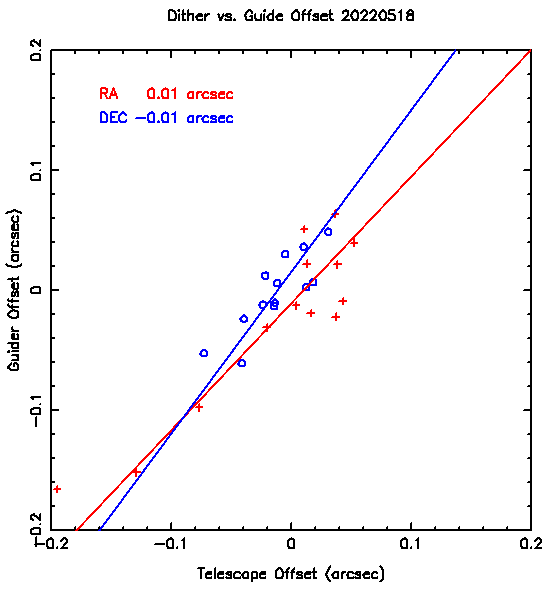}
    \caption{Comparison of telescope field center offsets, determined for
each exposure in a  dither sequence,
to offsets measured by guide stars at the same time.}
\label{fig:dither}
\end{figure}

\begin{figure}[ht]
    \centering
    \includegraphics[width=4in]{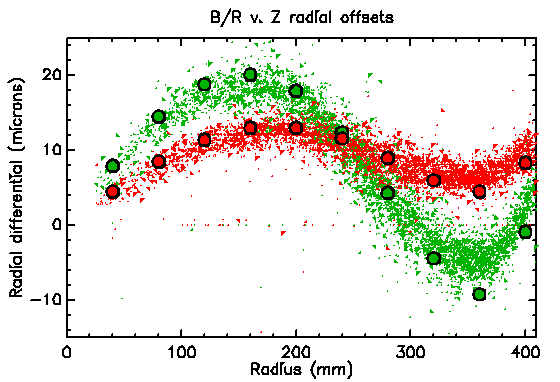}
    \caption{Radial distortion (in microns)
in B (green points) and R (red points)
channels relative to the Z channel as measured by the dither tests.
Open circles are the optical model
predictions.}
\label{fig:chromatic}
\end{figure}

A couple of cross-checks can be done to demonstrate that the dither tests
are functioning properly.  First, the dither analysis determines, among other
things, the offset in the field center from one exposure to the next.
These offsets should be reflected in the guide star offsets measured using
the guide CCDs at the
same time.  Figure \ref{fig:dither}
shows such a comparison, demonstrating that the
mean field offset does, indeed, match the guide star offset.  Additionally,
there is no systematic offset between the two measures, meaning that
the field is properly centered on the fibers if the guider error is
zero.  (This result also demonstrates that there is no remaining systematic
offset in the GIF metrology system.)
Second,
the corrector is not perfectly achromatic, so the radial variation in
positioner offsets among the three channels should match that predicted by
the optical model.  Figure \ref{fig:chromatic}
shows this comparison as well.  

\section{Off-Sky Operation}\label{sec:offsky}

The FVC is used frequently to image the focal plane without being on-sky,
most commonly to study positioner performance.  PM is used to analyze
these images and, depending on what tests are being done, supports
multiple modes of operation.

\subsection{Initial calibration}

PM normally relies on spotmatch to identify and measure the locations of
fiducials, but doing so relies on having an approximate mapping of FVC
pixels to the focal plane, which is not known initially.  Thus, PM has
its own code to identify the spot patterns of fiducials.  A single FVC
image of the back-illuminated fiducials is obtained.  5 of the 12
fiducials on each petal are located along an arc near the edge of the focal
plane, and these can be distinguished from the others. One petal has
an extra fiducial along this arc for symmetry breaking.  This extra fiducial
is identified with an automated algorithm,
and once done, the initial mapping to the focal
plane can be determined in a straightforward fashion.

\subsection{Petal Offsets}
\label{sec:petal}

The focal plane is assembled from 10 individual petals, and, when these are
assembled, there are small offsets that are not known in advance.  A set
of 10 FVC images of the backlit fiducials is obtained, and a solution is
obtained for the focal plane distortion with three additional calibration
coefficients enabled for each petal - two for $x,y$
offsets and one for rotation
about the nose of the petal.  A global constraint is imposed that the
average of all offsets and rotation be zero.  This process is operationally
how the CS5 system is defined.  Once these extra coefficients
are determined for each petal, they are kept frozen so as to reduce
the number of degrees of freedom in the measurement of distortions during
normal operations.

The rms residuals after fitting a distortion model to fiducials measured
in a typical FVC image are about 16$\mu$m rms 1-D.  These errors are presumably
due to errors in the laboratory metrology of the fiducial prior to petals
being delivered to the telescope.  One could average together many measurements
and improve the internal consistency of the petal metrology, although
this process has not been found to be essential yet.

\subsection{SPOTS for turbulence calibration}

The turbulence calibration relies on knowing the relative positions
of each fiducial or non-operational positioner, but their absolute location
in the focal plane is not needed (nor is it known a priori for the
positioners).  Instead, a series of FVC images are averaged to get the
coordinates empirically.  In practice, the coordinates will
also depend on the ADC settings and thus telescope position on the sky.
Further, there may be focal plane deformation due to gravity.  Therefore,
a program termed ``SPOTS" was created in which
a large number of FVC images is obtained over a range of hour
angles and declinations.  The pixel coordinates of each fiducial
and positioner fiber are transformed to the
focal plane using the as-built optical model, and each coordinate
is modeled as a constant plus a polynomial
dependence on hour angle and declination; the data are used to constrain
the coefficients of this model.

\subsection{Laboratory Tests}

Two spare petals were set up for conducting tests at LBL in a clean room.
A separate FVC was provided for these tests.  Several modifications were
needed to PM to map FVC images to each petal.

\begin{itemize}
\item There is no optical corrector, so a set of files to define an
optical model with no power was created.
\item The focal surface of the petals is an asphere and is viewed at a finite
distance, which means that there
is quite a bit of ``distortion'' in mapping FVC pixels to the focal surface.
\item The petals and FVC could be moved about, so procedures were developed
to redetermine basic calibration constants (position angle and demagnification)
more easily.
\end{itemize}

\section{Focal Plane Stability}\label{sec:stability}

PM generates a large amount of diagnostic information that can be used
to look for trends or uncover problems in the hardware.  The following are
a few examples.

\begin{enumerate}
\item The relative astrometric locations of the GFAs are measured for each
acquisition image, and long-term trends are monitored.  A simple diagnostic
involves monitoring the rms errors in astrometric solutions for fields near
zenith for degradation over time.  No such degradation has been seen thus
far.

\item The rms errors in astrometric calibration of the FVC images are
measured at the same time.  No significant trends have been noticed.

\item The rms errors in FVC to focal plane mapping are tracked.  These have
been stable over time.

\item The rms errors in the GIF to focal plane calibration are tracked.
These have been stable over time.

\item The dependence of various calibration coefficients
vs.\ telescope temperature are
tracked.  As an example, Fig.\ \ref{fig:fvctemp} shows the dependence
of the FVC scale factor on telescope temperature.  The slope
is about $-2.2\times 10^{-5}$ K$^{-1}$.  This value is
several times greater than the temperature coefficient of either
silicon or BK7 glass (both in physical dimension and refractive index)
and likely arises from thermal expansion of
the aluminum structure holding the FVC and lens (without refocusing).
By contrast, the
GFA astrometric scale factor shows no particular dependence on telescope
temperature, which is consistent with the focal plane being thermally
controlled (and kept in focus).

\begin{figure}[ht]
    \centering
    \includegraphics[width=4in]{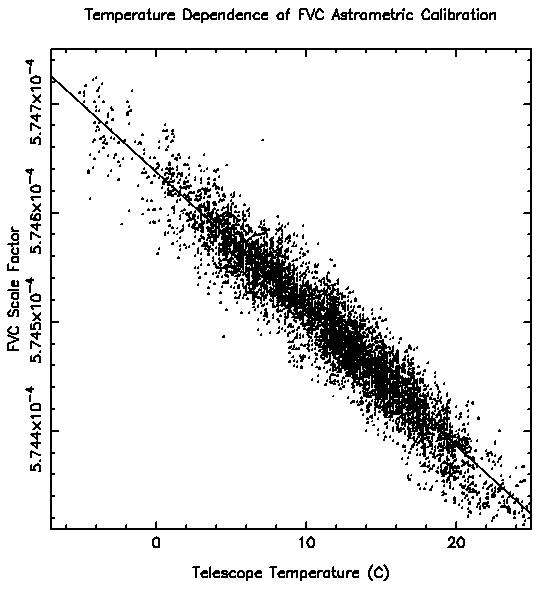}
    \caption{Temperature dependence of the FVC astrometric scale
factor.  Scale factor is in units of angular degrees per FVC pixel.}
\label{fig:fvctemp}
\end{figure}

\end{enumerate}

\section{Conclusions and Lessons Learned}\label{sec:conclusions}

The DESI
astrometric calibration systems and processes have demonstrated that
fiber positioning can be achieved repeatably and
reliably with an accuracy of 8$\mu$m 1-D (11$\mu$m 2-D), which meets the DESI
requirement for lateral positioning error.

Future large spectroscopic surveys currently being discussed
may possibly make use of a fiber view camera system of some sort, and
the DESI experience may provide some guidance as to their design and
implementation.  In case it might help, the following are some lessons learned
from the DESI experience:
\begin{itemize}
\item Ensure that the end-to-end astrometric calibration process is fully
defined at an early stage.
In DESI, initial planning focused on mapping distortion in the
corrector optics using the FVC along with the fiducials, but not enough
attention was paid to mapping from the focal plane to the sky or the
FVC to the sky.  The initial error budget omitted several
contributions that were important while including other
contributions that were unimportant.  Requirements on the performance of
the optical corrector did not account for the fact that the FVC images
only the chief ray from a back-illumnated fiber, which is affected most
by high frequency components of the polishing error budget.

\item Provide a direct mechanism to connect the focal plane to the sky.
In the case of the Sloan Digital Sky Survey \citep{owen98},
this connection was accomplished
by using coherent fiber bundles that were positioned in holes drilled
in a plug plate identical to the holes used for positioning
target fibers.  These bundles were distributed throughout the focal plane.
In the case of the 2dF instrument at the Australian Astronomical Telescope
\citep{lewis02}, a focal plane imaging system is used, including
a type of coherent imager that utilizes a cluster of fibers,
but only 4 are available and are normally placed at the edge of the field.
Coherent imagers require different routing for their fibers than for
the target fibers, which would have been difficult for DESI and accounts for
why they were not used.

\item Include a sufficient number of fiducials in the focal plane to
allow for monitoring and correction of turbulence.  DESI started with
only 113 such fiducials, whereas of order 300 or more are needed.

\end{itemize}

\section{Acknowledgements}

The authors would like to thank Michael Lampton, Tim Miller, and Charlie
Baltay for many useful discussions.  The DESI
collaboration and the authors in particular
are indebted to the late Michael Lampton for his
many contributions to the success of this project and mourn
his passing.  

This work was produced, in part, by Fermi Research Alliance, LLC under
Contract No. DE-AC02-07CH11359 with the U.S. Department of Energy. 

This material is based upon work supported by the U.S. Department of Energy
(DOE), Office of Science, Office of High-Energy Physics, under Contract No.
DE-AC02-05CH11231, and by the National Energy Research Scientific Computing
Center, a DOE Office of Science User Facility under the same contract.
Additional support for DESI was provided by the U.S. National Science
Foundation (NSF), Division of Astronomical Sciences under Contract
No.\ AST-0950945 to the NSF's National Optical-Infrared Astronomy Research
Laboratory; the Science and Technology Facilities Council of the United
Kingdom; the Gordon and Betty Moore Foundation; the Heising-Simons
Foundation; the French Alternative Energies and Atomic Energy Commission
(CEA); the National Council of Science and Technology of Mexico (CONACYT);
the Ministry of Science and Innovation of Spain (MICINN), and by the DESI
Member Institutions: \url{https://www.desi.lbl.gov/collaborating-institutions}.
Any opinions, findings, and conclusions or recommendations expressed in this
material are those of the author(s) and do not necessarily reflect the
views of the U. S. National Science Foundation, the U. S. Department of
Energy, or any of the listed funding agencies.

The authors are honored to be permitted to conduct scientific research on
Iolkam Du'ag (Kitt Peak), a mountain with particular significance to the
Tohono O’odham Nation.

For more information, visit \url{https://www.desi.lbl.gov/}.

This work has made use of data from the European Space Agency (ESA) mission
Gaia (\url{https://www.cosmos.esa.int/gaia}), processed by the Gaia Data
Processing and Analysis Consortium
(DPAC, \url{https://www.cosmos.esa.int/web/gaia/dpac/consortium}).
Funding for the DPAC has been provided by national institutions,
in particular the
institutions participating in the Gaia Multilateral Agreement.

\section{DATA AVAILABILITY}

All data points shown in the published graphs are available in a
machine-readable form at {https://doi.org/10.5281/zenodo.8302266}.

\end{document}